\documentclass[aps,preprint,superscriptaddress]{revtex4}

\usepackage{amsmath}
\usepackage{amssymb}
\usepackage{graphicx}

\newcommand{\diff}{\text{d}}

\newcommand{\mean}[1]{\langle #1 \rangle}
\newcommand{\Mone}{{$\mathcal M1$}}
\newcommand{\Mtwo}{{$\mathcal M2$}}
\newcommand{\Mthree}{{$\mathcal M3$}}

\newcommand{\Aone}{{$\mathcal A1$}}
\newcommand{\Atwo}{{$\mathcal A2$}}
\newcommand{\Athree}{{$\mathcal A3$}}

\def\fig#1{Fig.~\ref{#1}}

\begin{document}

\title{Dynamics of protein-protein encounter: a Langevin equation approach with reaction patches}

\author{Jakob Schluttig}
\affiliation{University of Heidelberg, Bioquant 0013, Im Neuenheimer Feld 267, 69120 Heidelberg, Germany}
\affiliation{University of Karlsruhe, Theoretical Biophysics Group, Kaiserstrasse 12, 76131 Karlsruhe, Germany}
\author{Denitsa Alamanova}
\affiliation{Center for Bioinformatics, Saarland University, 66041 Saarbr\"ucken, Germany}
\author{Volkhard Helms}
\affiliation{Center for Bioinformatics, Saarland University, 66041 Saarbr\"ucken, Germany}
\author{Ulrich S. Schwarz}
\email[E-mail: ]{ulrich.schwarz@bioquant.uni-heidelberg.de}
\affiliation{University of Heidelberg, Bioquant 0013, Im Neuenheimer Feld 267, 69120 Heidelberg, Germany}
\affiliation{University of Karlsruhe, Theoretical Biophysics Group, Kaiserstrasse 12, 76131 Karlsruhe, Germany}

\begin{abstract}
  We study the formation of protein-protein encounter complexes with a
  Langevin equation approach that considers direct, steric and thermal
  forces. As three model systems with distinctly different properties
  we consider the pairs barnase:barstar, cytochrome c:cytochrome c
  peroxidase and p53:MDM2. In each case, proteins are modeled either
  as spherical particles, as dipolar spheres or as collection of
  several small beads with one dipole. Spherical reaction patches are
  placed on the model proteins according to the known experimental
  structures of the protein complexes. In the computer simulations,
  concentration is varied by changing box size.  Encounter is defined
  as overlap of the reaction patches and the corresponding first
  passage times are recorded together with the number of unsuccessful
  contacts before encounter. We find that encounter frequency scales
  linearly with protein concentration, thus proving that our
  microscopic model results in a well-defined macroscopic encounter
  rate. The number of unsuccessful contacts before encounter decreases
  with increasing encounter rate and ranges from 20--9000. For all three models,
  encounter rates are obtained within one order of magnitude of the
  experimentally measured association rates.  Electrostatic steering
  enhances association up to 50-fold. If diffusional encounter is
  dominant (p53:MDM2) or similarly important as electrostatic steering
  (barnase:barstar), then encounter rate decreases with decreasing
  patch radius. More detailed modeling of protein shapes decreases
  encounter rates by 5--95 percent. Our study shows how generic
  principles of protein-protein association are modulated by molecular
  features of the systems under consideration. Moreover it allows us
  to assess different coarse-graining strategies for the future
  modelling of the dynamics of large protein complexes.
\end{abstract}

\date{\today}

\maketitle

\section{Introduction}

Protein-protein interactions play key roles in many cellular processes
such as signal transduction, bioenergetics, and the immune response
\cite{helms_08}. Moreover, many proteins function in the context of
protein complexes of variable sizes and lifetimes.  Examples of such
complexes are ribosomes, polymerases, spliceosomes, nuclear pore
complexes, cytoskeletal structures like the mitotic spindle or actin
stress fibers, adhesion contacts, the anaphase-promoting complex, and
the endocytotic complex \cite{alberts_02}.  For yeast, 800 different
core complexes have been identified, suggesting the existence of 3000
core complexes for humans \cite{gavin_06}. In addition it has been
shown for yeast that most protein complexes are assembled just-in-time
during the course of the cell cycle \cite{lichtenberg_05}. In fact many
protein complexes in the cell are highly dynamic, with fast turnover
of many components. One can argue that their dynamics,
although experimentally very hard to access, is biologically more
relevant than their equilibrium properties. Therefore a systematic
understanding of the dynamics of protein complexes in cells is one of
the grand challenges in quantitative biology.

The elementary unit of all of these cellular processes is the
bimolecular protein-protein interaction.  The strength and specificity
of protein-protein association are determined by the integrated effect
of different interactions, including shape complementarity, van der
Waals interactions, hydrogen bonding, electrostatic interactions and
hydrophobic effects. For example, the importance of electrostatic
interactions has been demonstrated by experimental measurement at
different ionic strengths \cite{schreiber_96}.  To a first
approximation, bimolecular reactions are characterized by on- and
off-rates. The equilibrium association constant (or affinity) then
follows as the ratio of the two. From a conceptual point of view, on-
and off-rates are very different.  On-rates are commonly believed to
be controlled by the diffusion properties as well as by long-ranged
electrostatic interactions, whereas off-rates are rather controlled by
short-ranged interactions like hydrogen bonding and van der Waals
forces.

The main features of the dynamics of protein association can be
conceptualized within the framework of the \emph{encounter complex}
\cite{schreiber_02}. To this end, the association is divided into two
parts. First, mutual entanglement - the encounter complex - is
achieved by the proteins due to a transport process including mainly
diffusion but also electrostatic steering on small length scales
\cite{spaar_06}. If diffusion-controlled, classical continuum
approaches can be used to describe this part of the process
\cite{berg_85}. To form the final complex, the system then has to
overcome a free energy barrier due to local effects like dehydration
of the binding interface \cite{ahmad_08}. Due to the various molecular
contributions involved in this step, here the two binding partners
essentially have to be modelled at atomic detail. Moreover the solvent
may need to be treated explicitly and one might has to account for
conformational changes \cite{lange_08}.  Thus, it appears reasonable
to use the encounter complex as a crossover point from a detailed,
atomistic treatment to a coarse-grained model and vice versa.

Thermal fluctuations are an essential element of protein-protein
encounter because they allow the two partners to exhaustively search
space for access to the binding interface. From the viewpoint of
stochastic dynamics, protein-protein association is a first passage
time problem which can be addressed mathematically in the framework of
Langevin equations.  The application of Langevin equations to
association phenomena goes back to early work in the colloidal
sciences \cite{smoluchowski_17, debye_42}.  In these early approaches,
the reactants were considered to have small spatial extensions and to
be uniformly reactive. For large biomolecules like proteins, the
situation is fundamentally different. Typically, proteins and other
biomolecules have specific sites on their surface, where a particular
binding reaction can take place. Therefore, such binding events are
subject to intrinsic geometric constraints for every particular
protein-protein pair or larger assembly. The standard model for
ligand-receptor interaction was introduced by Berg and Purcell in the
context of chemoreception based on the idea of using reactive patches
to model anisotropic reactivity \cite{berg_77}. Due to anisotropic
reactivity, also rotational diffusion becomes important.  Shoup et
al.\ showed that the effect of rotational diffusion can strongly
increase the association rate between a receptor with a flat reactive
patch and uniformly reactive ligands \cite{shoup_81}. Later, analytic
expressions for the association rate between two spherical particles
with both carrying a flat axially symmetric \cite{zhou_93} and
asymmetric \cite{schlosshauer_02, schlosshauer_04} reactive patch were
derived.  Similar concepts were also applied by Schulten and
colleagues \cite{szabo_80}.

For many important aspects, analytical approaches are not possible and
computer simulations are required. This approach has been used early
for protein-protein association \cite{northrup_88}. The importance of
electrostatic interactions for long ranged attraction was also
emphasized by Brownian dynamics simulations of protein-protein
encounter \cite{northrup_88, zhou_97, wade_98, elcock_99}. If atomic
structure is taken into account, then successful encounters are
defined by simultaneous fulfillment of two to three distance
conditions between opposing residues on the two surfaces
\cite{gabdoulline_97}. Brownian dynamics have also been used for the
simulation of high density solutions, e.g. by Bicout and Field who
studied a cellular ``soup'' containing ribosomes, proteins and tRNA
molecules \cite{bicout_96}, or by Elcock and coworkers who
simulated a crowded cytosol for 10$\mu$s \cite{mcguffee_06}.

In order to develop a quantitative framework for modelling the
dynamics of protein complexes, it is essential to understand the
relative importance of generic principles and molecularly determined
features of specific systems of interest.  Only a good understanding
of this issues will allow us in the future to develop reasonable
coarse-graining strategies to address also large complexes of biological
relevance.  In this study, we therefore address how general principles
guiding the diffusional association of biomolecular pairs are
modulated by their particular physicochemical properties.  To this end
we have selected three molecular systems of interest with different
steric and electrostatic properties. One of the best studied
bimolecular complexes is the extracellular ribonuclease Barnase and
its intracellular inhibitor Barstar. Both proteins carry a net charge
of $2e$ and $-6e$, respectively, which leads to a considerable
electrostatic steering \cite{chong_98, selzer_99, sheinerman_02,
  dong_03, wang_wade_04}.  Considering the structure of the two
proteins, Barnase has a bean-like form, matching well on a large
reactive area with the nearly spherical Barstar. A classic example of
electrostatically-driven protein association is the iso-1-cytochrome
{\it{c}} - cytochrome c peroxidase (Cytc:CCP) complex, charged with
$6e$ and $-13e$, respectively, and exhibiting dipoles aligned well
with the reactive areas \cite{northrup_88, pelletier_92}. Finally, we
selected the medically important complex of a peptide fragment of p53
and its inhibitor MDM2, which is used for anticancer drug
design. In this system, electrostatic attraction plays a minor role.
On the other hand, the steric match of the two surfaces is of
particular importance here. It is a perfect example of a key-lock binding
interface, where p53 is buried deep into a cleft on the MDM2 surface.

In this paper we systematically explore the effect of various
coarse-graining procedures on the rate for protein-protein encounter
for the three selected model systems. We revisit early approaches
based on Langevin equations and combine them with current knowledge on
molecular structure. The paper is organized as follows: In Sect.\
\ref{model_and_methods}, we present our different stochastic models
and describe the methods we use to parameterize the three considered
bimolecular model systems. Sect.\ \ref{results} contains the main
findings of our study, which are discussed and summarized in Sect.\
\ref{summary_and_discussion}.

\section{Models and Methods \label{model_and_methods}}

\subsection{Modelling proteins at different levels of detail}

One aim of this work is to determine how important specific details of
the model proteins are with respect to the association properties.
Therefore, we considered three different levels of detail as depicted
in Fig.\ \ref{different_model_scheme} for the three chosen systems.
In the most generic approach (\Mone), we only considered the steric
interaction between spherical particles covered with reaction patches.
As a first refinement (\Mtwo), an effective Coulombic interaction was
introduced using the dipolar sphere model (DSM). Finally, since our
Langevin equation approach is particularly suited to capture
anisotropic transport, we consider a more refined version for protein
sterics (\Mthree). In this approach the excluded volume of each
protein was modeled by 8-25 smaller beads. \Mthree\ uses the DSM as
well. In Fig.\ \ref{different_model_scheme}, we also show the full
structures in the bottom row as surface representations, including the
locations of the binding interfaces.  In the following, the general
properties of the simulation model and the different techniques used
in this work will be explained.

\subsection{Diffusion properties}

The diffusion of the protein model particles is described by an
anisotropic $6 \times 6$ diffusion matrix in all versions of our model. In
Ref.\ \cite{torre_00}, de la Torre and coworkers present a method to
calculate this diffusion matrix from the pdb structure of a protein.
This method has been implemented in a software called HYDROPRO which
is provided online by the same authors ({\tt
  http://leonardo.fcu.um.es/macromol}). The basic concept is to put
spheres of a certain size at the position of any non-H atom. The
volume of these spheres effectively models a fixed hydration shell.
This construct is then filled up with smaller, densely packed, but
non-overlapping spheres. Since the hydrodynamic properties of a rigid
body are determined by its outer boundary only, a shell of these small
spheres is generated by deleting all spheres which have a maximum
number of possible neighbors.  To this shell a sophisticated technique
is applied, which has been developed by de la Torre and colleagues
over the years, to calculate the diffusion matrix of such a cluster of
non-overlapping spheres (see references in \cite{torre_00}). Several
system properties are implicitely contained in the mobility matrix,
such as ambient temperature $T_a=293$K, as well as the density and
dynamic viscosity of the solvent, where we chose the respective
parameters of water, $\rho=1$g$/$cm$^3$ and
$\mu=10^{-3}$Pa s.  For simplicity, hydrodynamic
interactions were not introduced in our models, because the
corresponding effect on the association rates is expected to be well
below 10\% \cite{antosiewicz_96}.

\subsection{Langevin equation and simulation method}

For the integration of the Langevin equation, which describes the
stochastic motion of the particles, we follow an approach which has
been recently developed to model cell adhesion via reactive receptor
patches \cite{korn_06, korn_07}. Let ${\mathbf X}_t$ be a
six-dimensional vector describing position and orientation of a
particle at time $t$. Since the noise due to Brownian motion is
additive (which means that it does not depend on ${\mathbf X}_t$ due
to a constant mobility matrix $\mathsf M$), the Langevin equation is
given by:
\begin{align}
\partial_t {\mathbf X}_t = {\mathsf M}{\mathbf F}+{\mathbf g}_t \text{ .}
\end{align}
Here, $\mathbf F$ is a six-dimensional vector containing the force and torque
acting on the particle, and ${\mathbf g}_t$ denotes Gaussian white noise:
\begin{align}
\langle {\mathbf g}_t \rangle = 0 \text{ ,} \quad 
\langle {\mathbf g}_t {\mathbf g}_{t'} \rangle = 2 k_B T_a {\mathsf M}
\delta(t-t') \text{ .}
\end{align}
As explained in App.\ C of Ref.\ \cite{korn_07}, the Euler algorithm can be
used to solve a discretized version of this equation:
\begin{align}
{\mathbf X}(t+\Delta t)= {\mathbf X}(t) + \mathsf M{\mathbf F}(t)\Delta t +
 {\mathbf g}(\Delta t) + {\mathcal O}(\Delta t^2)\text{ .}
\label{langevin_equation_discrete}
\end{align}
For proteins, the typical orders of magnitude are $D=10^{-6}$cm$^2$s$^{-1}$
and $R=1$nm. Therefore, a reasonable choice for the time step is $\Delta
t=1$ps, as this leads to a mean step length of $\sqrt{D \Delta t}=0.01$nm.

The mobility matrix of a particle is defined in a particle-fixed coordinate
system. Thus, the whole step has to be calculated in terms of particle-fixed
coordinates and then transformed to the laboratory coordinate
space. In particular, this transformation implies a rotation $\mathsf R$
regarding the orientation of the particle. Special attention has to be
payed to the force $\mathbf F$, which is typically calculated in the
global frame of reference and hence has to be transformed to particle space
before Eq.\ \ref{langevin_equation_discrete} can be evaluated.  This
back-transformation is achieved by applying $\mathsf R^{-1}$ to $\mathbf F$.
Since rotation matrices $\mathsf R$ simply consist of a list of orthonormal
vectors, their inverse is equal to the transposed
matrix $\mathsf R^{-1}= \mathsf R^T$. Thus Eq.\
\ref{langevin_equation_discrete} can be rewritten as
\begin{equation}
{\mathbf X}(t+\Delta t)=
\quad {\mathbf X}(t) + \mathsf R\left[\mathsf M\left(\mathsf R^T{\mathbf F}(t)\right)\Delta t +
 {\mathbf g}(\Delta t)\right] + {\mathcal O}(\Delta t^2)\text{ .}
\label{langevin_equation_discrete_rotated}
\end{equation}
As $\mathbf F$ and $\mathbf g$ are six-dimensional and contain information
about torque and rotation, Eq.\ \ref{langevin_equation_discrete_rotated} is
only formally correct, as $\mathsf R$ acts on both the translational and
rotational parts of the respective vectors separately.

In each step of the simulation, a displacement vector $\Delta{\mathbf X}(t)$
is drawn for each particle as described above. If this global displacement
leads to any violation of the hardcore repulsion, all suggested displacements
are rejected and new $\Delta {\mathbf X}(t)$ are calculated. This procedure
continues, until an update of all positions and orientations is found which
does not lead to any overlap.  In this way, the constraint according to the
excluded volume effect is included in the stochastic motion.  The spherical
reactive patches are not taken into account for the steric interactions, i.e.\
they may not only overlap pairwise but also with the model particles.  One
would expect that our procedure leads to errors of order $\Delta t$ if two
particles are in close proximity of order $\sqrt{D \Delta t}$.
However, it has been shown for a different system \cite{ramanathan_07} that
in practise the deviation from the expected behavior is very small and thus the
approach is reasonable.

\subsection{Anisotropic versus isotropic diffusion}

As mentioned before, the $6 \times 6$ mobility matrix $\mathsf M$ represents
anisotropic diffusion. For large times, anisotropic diffusion crosses
over into isotropic diffusion because the information about the
initial orientation gets lost after a certain relaxation time due to
the rotational diffusion \cite{han_06}. In general, translational and
rotational diffusion are coupled so that large time steps cannot be
used. However, for the particular systems studied here, we found that
the diffusive coupling is a very small effect. In particular, the
major entries in the diffusion matrix of the proteins used here
according to HYDROPRO multiplied with different powers of the Stokes
radius $R\sim10^{-7}$cm to make the dimensions comparable are
$D_\text{tt}/R^2\sim10^{8}\text{s}^{-1}$,
$D_\text{rr}\sim10^{7}\text{s}^{-1}$,
$D_\text{tr}/R\sim10^{5}\text{s}^{-1}$.  Therefore, the effect of
diffusive coupling is $10^{-2}$ and $10^{-3}$ smaller than rotational
and translational diffusion, respectively.  Finally, the typical time
scale at which the cross-over is expected can be calculated to be
$1/(6D_\text{rr})\approx10$ns. Time steps of this magnitude were
rarely used in the simulations (see below), so that for most of the
steps, the anisotropicity is well preserved. Therefore we can safely
neglect changes in the anisotropicity of the mobility matrix.

\subsection{System size and time step adaption}

The simulations were performed in a cubic box with periodic boundary
conditions.  Schreiber and Fersht used concentrations between
$0.125\mu$M and $0.5\mu$M in their experimental studies of the
association rate of the Barnase:Barstar complex \cite{schreiber_96}.
The average volume containing one particle at a concentration $c$ is
$1/cN_A$ with the Avogadro number $N_A=6\cdot10^{23}$mol$^{-1}$.
Hence, the edge length of a cubic boundary box representing
concentration $c$ can be calculated from $L=\sqrt[3]{V}=1/\sqrt[3]{c
  N_A}$. E.g.\ $c=0.125\mu$M leads to $L\approx2370$\AA\ for one pair
of particles, which is two orders of magnitude larger than the size of
the proteins. Due to this low density, the first passage times (FPT)
for encounter can be expected to be much longer than the chosen time
step. For computational efficiency, we therefore used a variable time
step in our simulations.  Van Zon and ten Wolde suggested a method to
avoid unwanted collisions when they introduced their Green's function
reaction dynamics (GFRD) \cite{wolde_05_a}. In contrast to our work,
however, this method is based on isotropic diffusion.  Generalizing
the GFRD to anisotropic diffusion is out of the scope of our work and
we therefore used the following scheme. We first note that in GFRD
each time step is chosen such that it includes the next reaction. In
our case, we also want to investigate the stochastic dynamics before
the next encounter event takes place. Thus a large time step is not
chosen to include the next encounter, but to bring the system to such
a configuration that encounter becomes more likely.  This step can be
well represented by isotropic diffusion with an overall diffusion
constant $D = (\mathcal D_{11}+\mathcal D_{22}+\mathcal D_{33})/3$
following from the anisotropic diffusion matrix.

For an isotropic random walk, the displacement probability is given by
a Gaussian distribution with spherical symmetry. Thus, large spatial
steps are exponentially suppressed, which makes a step of size $\Delta
r^H_\text{max}\ge H\sqrt{6D\Delta t}$ an $H$-sigma event. By setting
$\Delta r^H_\text{max}=\mathsf{min}\{r_{ij}^\text{eff}\}$ the smallest
effective particle distance in the system, where effective means the
distance of the surfaces $r_{ij}^\text{eff}=|\mathbf r_i-\mathbf
r_j|-R_i-R_j$ with $R_{i}$ determining the maximal steric interaction
radius of particle $i$, one can estimate a reasonable time step for
which a collision is highly improbable. Van Zon and ten Wolde found
that the choice $H=3$ provides good results, combined with the fact
that wrongly sampling a collision event would need a certain direction
of the displacement in addition to the length. As the particles reach
close proximity $\mathsf{min}\{r_{ij}^\text{eff}\}\rightarrow0$, the
estimated $\Delta t$ vanishes and thus the simulation would be slowed
down infinitely.  Therefore, there has to be some lower boundary for
the time step $\Delta t_\text{min}$, which is generally chosen to be
$\Delta t_\text{min}=1$ps as explained earlier. Thus, the adapted time
step is given by:
\begin{align}
  \Delta
  t_\text{ad}=\mathsf{min}\left\{\dfrac{H^2}{6D}\left(\mathsf{min}\{r_{ij}^\text{eff}\}\right)^2
  ; \Delta t_\text{min}\right\} \text{ .}
\end{align}
In practice, most time steps are in the ps-range, with very few time
steps coming up to the ns-range.

\subsection{Electrostatic interactions}

Electrostatic interactions are known to play an important role in protein
association. To study the effect of electrostatics in our generic model, the
models \Mtwo\ and \Mthree\ utilized the dipolar sphere model (DSM), following
Refs.\ \cite{eltis_91, gorba_04}. The DSM effectively models a monopole and
dipole interaction by summing over the interactions of three charges, one
positioned in the center of each particle, and two close to its surface in
opposite positions. Taking into account the Debye screening function due to
the presence of counter ions in solution, the electrostatic interaction energy
between two charges $q_{i/j}$ at positions $\mathbf r_{i/j}$ respectively with
distance $r_{ij}=|\mathbf r_{ij}|=|\mathbf r_j-\mathbf r_i|$ is:
\begin{align}
W_{ij}=\frac{1}{4\pi\varepsilon_0\varepsilon_r} q_i
q_j \frac{e^{-\kappa(r_{ij}-B_{ij})}}{(1+\kappa B_{ij})r_{ij}} \text{ .}
\end{align}
Here, $\kappa=l_D^{-1}$ is the inverse Debye screening length, which
typically has a value of $\approx1$nm under physiological conditions.
We assume a value of $\varepsilon_r=78$ for the relative static
permittivity of the medium, which reflects the properties of water at
ambient temperature.  $B_{ij}$ is a correction to the screening of
charges which are placed in an object like a protein which has no free
charges inside. Taking $b_{i/j}$ as the closest distance of $q_{i/j}$
from the surface of the surrounding protein, it is approximately given
by $B_{ij}=b_i+b_j$. This potential leads to a force of charge $q_j$
on $q_i$:
\begin{align}
\mathbf F_{ij}&=-\nabla_{\mathbf r_i} W_{ij}=
-\frac{\partial  W_{ij}}{\partial r_{ij}}\cdot\left( \nabla_{\mathbf
  r_i}r_{ij}\right) \notag \\
&=-\frac{1}{4\pi\varepsilon_0\varepsilon_r}
 q_i q_j \frac{e^{-\kappa(r_{ij}-B_{ij})}(1+\kappa r_{ij})}{(1+\kappa
   B_{ij})r_{ij}^2}\frac{\mathbf r_{ij}}{r_{ij}}
 \text{ .}
\end{align}
As our simulation uses periodic boundary conditions, actually an infinite
number of copies exists for every charge. However, due to the very fast decay
of the screened electrostatic interaction, only the minimum image distance of two
charges is considered in the force calculation. Two model particles $m$ and
$n$ feel the sum of the Coulomb forces $\mathbf F_{ij}$ between all pairs of
the three complementary charges mimicking the monopolar and dipolar
interactions. Thus the full force between particle $m$ and $n$ is $\mathbf
F_{mn}=\sum_{i=1}^3\sum_{j=1}^3 \mathbf F_{ij}$, where $i$/$j$ run over the
charges of $m$/$n$ respectively.

As explained earlier, the action of the force on a particle in the
Langevin equation is weighted with the mobility matrix $\mathsf M$.
The HYDROPRO software directly gives the diffusion matrix $\mathcal
D=k_BT_a\mathsf M$. This means that in our case the force action
should be rewritten as $\mathsf M\mathbf F\Delta t$=$\mathcal D\mathbf
F\Delta t/k_BT_a$.  Considering a time step of $\Delta t=10$ps and a
typical diffusion constant $D=10^{-10}\text{m}^2/\text{s}$, we have
$D|\mathbf F(1\text{nm})|\Delta t/k_BT_a\sim 10^{-13}$m and $D|\mathbf
F(4\text{nm})|\Delta t/k_BT_a\sim 10^{-15}$m for typical distances
$r_{ij}=1$nm and $r_{ij}=4$nm, respectively. In contrast, the typical
step length due to the Brownian motion is $\sqrt{D\Delta t}\sim
10^{-11}$m. This shows that the magnitude of electrostatic
interactions at distances of $1$nm is much smaller than thermal
energy. Therefore the effect of force is also not considered in our
adaptive scheme for the time steps. However, it can be expected that the
\emph{systematic} drift, albeit small, will still lead to an altered
encounter behavior.

\subsection{Parameterization}

Gabdoulline and Wade \cite{gabdoulline_97} used several criteria to
define contact areas of bimolecular protein complexes. In our studies,
we define the contact area to consist of those atoms in the two
interacting proteins that are at $5$\AA\ or less distance from an atom
of the complementary protein. The center of mass of these atoms is
considered as the center of the reactive area. For \Mone\ and \Mtwo,
the reactive patch is centered at the surface of the sphere modeling
the excluded volume such that it has the same relative direction from
the center of mass as obtained by the method described. In the case of
\Mthree, the center of the patch is set to the center of the reactive
area.

The contact area has a diameter of approximately 10\AA\ to 20\AA\ for
the three systems studied here. Following earlier Brownian dynamics
simulations with atomistic details \cite{spaar_06} we have performed
an in-depth analysis of the free energy landscape and the encounter
state of the protein complexes considered in this work (unpublished
results). This showed that the encounter complex is typically located
at relative separations of the two protein surfaces of about 10\AA\
compared to their positions in the final complex. As the spherical
reactive patches used in this study simultaneously determine both the
size of the contact area on the surface and the distance above their
surface at which an encounter will be possible, values in the range of
5\AA\ to 10\AA\ seem to be reasonable. Note that as long as physical
considerations do not dictate non-spherical reaction patches, the
spherical choice is highly favorable for computational efficiency.

As already stated in the beginning, two types of excluded volume structures
are taken into account. In the first case, used in \Mone\ and \Mtwo, the
proteins are assumed to have an approximately spherical form. The radius for
the model spheres determining the hard core interaction follows as the
radius of gyration of the protein, which is also calculated by the HYDROPRO
software. The underlying data in the more detailed approach \Mthree\ is
obtained using the AtoB bead modeling software \cite{byron_97, rai_05}. In
this way, the three-dimensional structure of the proteins is modeled with a
comparably small number of 8 to 25 spheres of different sizes.

The monopole charge is the sum of all elementary charges in a protein
and is placed at the center of the respective model particle.  The
dipole moment $\mathbf p$ is obtained by summing over the product of
all atomic charges due to the xyz force field and their relative
position to the center of mass. In the model, it is represented by two
opposing charges which are positioned along the direction of $\mathbf
p$ and at a distance $r_p=R_\text{gyr}-4$\AA. The magnitude $p'$ is
chosen such that $|\mathbf p|=2p'r_p$. We found that the particular
choice of $r_p$ does not have a noticeable influence on the results.
The resulting parameterization is given in Tab.\ \ref{parameter_table}
for the proteins considered here.

\section{Results \label{results}}

\subsection{Encounter frequency and encounter rate}

Langevin dynamics simulations were performed for cubic boxes
containing two model proteins. Simulations were conducted until the
encounter condition was met for the first time (typically after
milliseconds). Because in our Langevin simulations we measure the
distribution of first passage times (FPT) to encounter, from which we
can deduce the mean first passage time (MFPT) $\mean{T}$, the
encounter frequency is defined as $k=1/\mean{T}$. This choice is
motivated by the fact that for a Poisson-like process, the
distribution of first passage times is given by $f(T)=k e^{-kT}$ and
therefore the encounter frequency $k$ indeed satisfies $k=1/\mean{T}$. As
the preparation of a comparable experiment would never allow knowing
the particular initial positions and orientations of the unbound
proteins, it makes sense to average over the possible initial
configurations in the computer simulations.  Therefore, we started a
large number of runs (typically $10^4$ to $10^5$) with random initial
positions and orientations for all involved model particles, under the
constraint that the initial pairwise distance is at least large enough
to prevent an immediate encounter. The ``first passage'' is defined as
the first overlap of two complementary reactive patches.
Interestingly, due to this averaging the first passage process becomes
Poisson-like, see Fig.\ \ref{randinit_histogram}. The data show a
clear exponential behavior.  This means that it is justified to use
the notion of an ``encounter frequency'', as the FPT distribution is indeed
represented by a single stochastic rate. The finite probability at
small FPT is due to the possibility that the two model particles are
already in close proximity when the simulation is started. The large
errors in the histogram at $T\rightarrow 0$ are caused by the fact
that exponentially sized histogram bins were used to sample the
behavior for small $T$. Therefore, events hitting a particular bin are
rare because of the small width of the bins at $T\rightarrow0$, which
then leads to bad statistics in this domain.

As the encounter process is purely diffusion limited in \Mone, one
would expect the encounter frequency to scale linearly with
concentration. Fig.\ \ref{randinit_concentrations} demonstrates for
the Barnase:Barstar system that this is indeed the case.  Hence, it is
reasonable to always scale the encounter frequencies with the inverse
concentration, as will be done for the rest of this work. We will
denote these rescaled quantities as \emph{encounter rate}, i.e.\ the
encounter rates have the dimension M$^{-1}$s$^{-1}$. In summary, we
have demonstrated here that our microscopic model leads to a
well-defined macroscopic encounter rate.

\subsection{Finite size effects}

In most of the simulations, only one instance of the final complex was
considered, i.e.\ one model particle of each kind. Using such small
systems could lead to undesired finite size effects. We therefore
considered the effect of having many particles in the simulation box.
Fig.\ \ref{randinit_finite_size} shows the simulation results for the
encounter frequency $k$ for an increasing number of Barnase:Barstar
pairs, while keeping the size of the boundary box constant. In order
to understand the expected effects, consider a system with molecules
of Barnase ($\mathsf A$ and $\mathsf A'$) and two molecules of Barstar
($\mathsf B$ and $\mathsf B'$) randomly distributed over the boundary
box. The relative alignment of any pair of $\mathsf A$s and $\mathsf
B$s is therefore random again. For a particular pair the distribution
of times to first encounter will thus look very similar to the case
with a single pair in the box, which is a simple exponential decay
with respect to the encounter frequency $k_1$: $f_1(T)=k_1 \exp[-k_1
T]$.  The probability that, e.g., the particular pair $\mathsf
A-\mathsf B$ reaches encounter at a certain time $t$ before the three
other possible pairs ($\mathsf A'-\mathsf B$, $\mathsf A-\mathsf B'$,
$\mathsf A'-\mathsf B'$), is therefore:
\begin{equation}
p(t)=\int\limits_0^\infty\diff t_1 \int\limits_{t_1}^\infty\diff t_2 
\int\limits_{t_1}^\infty\diff t_3 \int\limits_{t_1}^\infty\diff t_4 \,
 \delta(t_1-t)\prod\limits_{i=1}^4 k_1 e^{-k_1 t_i} =k_1 e^{-4 k_1 t} \text{ .}
\end{equation}
Thus, the probability that any of the four possible particle pairs
reaches encounter before the respective three other pairs do, is
$4\times p(t)$ as just calculated, i.e.\ $f_2(T)$ has again a Poisson
form like $f_1(T)$ and $k_2=4 k_1$. In general,
for higher numbers of particle pairs $N$, we expect to again find an
exponential distribution of the time to first encounter with the
encounter frequency $k_N=N^2 k_1$. This
quadratic behavior is nicely confirmed by the data shown in Fig.\
\ref{randinit_finite_size}, which suggests that even for small systems
with only two particles, no severe finite size effects have to be
expected. In particular, this rules out that larger numbers of
particles lead to noticeable three-body interactions or hindering of
the encounter process.

\subsection{Alignment during encounter}

One feature of special interest which we can address with our Langevin
equation approach is the pathway through which the encounter is
formed. We dissected the encounter process into several parts as
visualized in Fig.\ \ref{alignment_states_scheme}. At the start of
each run, the systems were prepared in the unaligned state \Aone, as
described earlier. A state of close approach which however does not
allow for binding is called \Atwo. The two model proteins will switch
between states \Aone\ and \Atwo\ a number of times $N$, until they
finally reach the encounter complex \Athree\ due to a favorable
combination of translational and rotational diffusion. In the
following, each occurance of \Atwo\ will be termed a \emph{contact}.
Thus $N$ counts the number of unsuccessful contacts before the
encounter is finally formed. A separate set of simulations was
performed to measure the distribution of $N$. Furthermore we analyzed
the distribution of return times $T_\text{off}$.  This is the time it
takes for two model proteins to get into contact again
(\Aone$\rightarrow$\Atwo) after having lost translational alignment
(\Atwo$\rightarrow$\Aone), i.e.\ after they were in close proximity.
Finally, we determined the distribution of resting times $T_\text{on}$
in translational alignment \Atwo\ before the two model particles
separated again.

As an example, Fig.\ \ref{alig_no_example} shows the distribution of
$N$ for the Barnase:Barstar model system at $c=0.5\mu$M in the
framework of \Mone.  Surprisingly the distribution of the number of
contacts has again a Poisson form. Note that the number of
unsuccessful trials in state \Atwo\ can be rather large (up to
$10^4$). We also found that the distribution of $N$ is roughly
independent of concentration.  This is reasonable, as after the two
proteins were in contact once, the further encounter process is guided
by returns to state \Atwo\ and thus should be more or less independent
of system size.

Fig.\ \ref{alig_onoff_example} shows that the return time
$T_\text{off}$ (plotted with the plus-symbol) is not exponentially
distributed. Instead, it follows a power law $p(T_\text{off})\sim
T_\text{off}^{-3/2}$ and undergoes an exponential cutoff due to the
finite size of the boundary box at large $T_\text{off}$. Therefore,
there is a high probability for very small return times, i.e.\
situations, where the two model proteins do not really separate, but
immediately after loosing translational alignment
(\Atwo$\rightarrow$\Aone) get closer again (\Aone$\rightarrow$\Atwo).
%Thus, an increase in $\mean{N}$ will not necessarily
%lead to a proportional increase of the time to encounter. 
The power law behavior of the return time is consistent with the
problem of a random walk to an absorber in three dimensions
\cite{polya_21,hughes_95}. In principal, these two situations are equivalent
since the relative motion of the two proteins while unaligned \Aone\
can be approximately understood as an isotropic random walk, and the
criterion for going over to translational alignment \Atwo\ reflects an
absorbing boundary in the configuration space of relative positions.

The distribution of resting times $T_\text{on}$ (plotted with the
cross-symbol in Fig.\ \ref{alig_onoff_example}) follows the same power
law as $f(T_\text{off})$, but the exponential cutoff occurs much
earlier.  The reason is that here the cutoff is determined by the
region in configuration space where the two model proteins are in
state \Atwo.  As this is much smaller than the whole volume of the
boundary box, in which they are unaligned and therefore in state
\Aone, a random walk in state \Atwo\ will end earlier.

The differences we obtain in the distributions of $T_\text{on}$ and
$T_\text{off}$ when using the variants \Mtwo\ and \Mthree\ compared to
\Mone\ are generally very small and unlikely to account for any
deviations in the overall encounter rates. Also, the distribution of
$N$ is always well described by a single exponential decay. However,
the inverse decay length $\mean{N}$ significantly varies between the
different situations. Therefore, changes in the overall encounter rate
are mainly caused by a different probability for reaching state
\Athree\ from state \Atwo. This is reasonable when considering that
the interactions are strongly localized and can thus only act
while the system is in the aligned state \Atwo.

\subsection{Three bimolecular systems with different physico-chemical interface properties}

So far we have only considered Barnase:Barstar ($\mathcal S1$) to
demonstrate how our computational model works. We now use our setup
for a more comprehensive investigation. In particular, we also apply
our method to two other systems, cytochrome c and its peroxidase
($\mathcal S2$) as well as the p53:MDM2 complex ($\mathcal S3$). Those
represent systems with different interface characteristics and where
the role of electrostatics is either much stronger ($\mathcal S2$) or
much weaker ($\mathcal S3$) than for $\mathcal S1$. To this end, all
the previously described quantities were measured for 8 different
concentrations $c=\{125,250,500,750,1250,2500,5000,7500\}$pM.
Furthermore, to find out how the choice of the radius of the reactive
patch affects the results, we used patch radii of $r=6$\AA\ and
$r=3$\AA\ in addition to the initially considered value of $r=10$\AA.

Tab.\ \ref{main_data_table} lists the encounter rates $k$ as obtained
from these simulations. The rates are all roughly of the same order of
magnitude.  Yet several interesting qualitative features are readily
apparent. First, for decreasing patch sizes, the rates generally
decrease. Second, this effect is weaker for \Mtwo\
compared to \Mone, which basically means that the electrostatic
attraction and orientation due to the dipole interaction are indeed
enhancing the encounter. The strongest effect of the electrostatic
interaction is obtained for Cytc:CCP, which is the system with the
largest monopole and dipole and the best alignment of the directions
of the dipoles and the reactive patches. On the other hand p53:MDM2 is
nearly unaffected by the effective charges, due to its weak monopole
charges and, additionally, an unfavorable alignment of the dipolar
interaction and the reactive surface area.  Furthermore, regarding the
results with detailed steric structure \Mthree, the effect on the rate
is correlated with the deviations of the protein forms from the
spherical excluded volume approach in \Mone\ and \Mtwo. This deviation
is smallest for Cytc:CCP and largest for p53:MDM2.

The findings for the encounter rate $k$ are also reflected in the
results for $\mean{N}$. As expected, an increase in $k$ correlates
with a decrease in $\mean{N}$. The only exception is Cytc:CCP observed
in \Mtwo, which is also special in regard to the effect of patch size.
Here, the effective Coulombic interaction is strongest and the dipole
moment is best aligned with the reactive patches. Therefore, having
reached state \Atwo\ once, the proteins do systematically orient
towards \Athree, while they are additionally strongly steered back
towards \Atwo\ when loosing their translational alignment. This
behavior is the stronger the closer the model proteins have approached
once -- i.e.\ for the case of small patch sizes, where state \Atwo\
implies the smallest distance. While this only explains the inversion
in the $\mean{N}$ behavior as a function of patch size $r$, $k$ is
obviously still slightly decreasing with smaller patch sizes. This can
be explained by the fact that the time to the first approach of state
\Atwo\ is larger for smaller patches, as this implies a smaller
relative distance. This obviously compensates the fact that afterwards
the encounter is formed even quicker, as reflected by the decreasing
$\mean{N}$.

The strong correlation between the encounter rate $k$ and the mean number of
contacts $\mean{N}$ is also evident from the correlation plot in Fig.\
\ref{correlation_plot}. Indeed, $k\sim\mean{N}^{-1}$ seems valid for
most of the different systems and models. It is noteworthy that
the prefactor is very similar in all cases. Basically, this means that one
unsuccessful contact takes the same amount of time on average, no matter what
the local details of the system are. This gets more obvious recalling the
distributions of the resting and return times $T_\text{on}$ and $T_\text{off}$
in Fig.\ \ref{alig_onoff_example}, which shows that
$\mean{T_\text{off}}>\mean{T_\text{on}}$. As the average time for one contact
will be approximately $\mean{T_\text{on}}+\mean{T_\text{off}}$, it is
dominated by $T_\text{off}$, which is only marginally influenced by the local
details of the system and the chosen model. Therefore it can be concluded,
that for $\mathcal S1$ and $\mathcal S3$ the incorporation of a more detailed
modeling approach influences $k$ and $\mean{N}$, but not the overall
characteristics of the encounter process.

The only exceptions for the clear correlation of $k$ and $\mean{N}$
are \Mtwo\ and \Mthree\ for the case Cytc:CCP ($\mathcal S2$), where
$k$ is nearly independent of $\mean{N}$ because of the strong
electrostatic interaction. This is consistent with the earlier
finding, that the behavior of Cytc:CCP is qualitatively different
\cite{northrup_88}, as its electrostatic interactions would facilitate
long-lived nonspecific encounters between the proteins that allowed
the severe orientational criteria for reaction to be overcome by
rotational diffusion. For all three systems studied, in \Mthree\ the
smallest patch size $r=3$\AA\ leads to a somewhat artificial slowing
down, because in this case an overlap of the patches is rather
hindered by the beads modeling the protein structure.

\subsection{Size of the reaction patches}

We next address the dependence of the data on the size of the reaction
patches in more detail. This behavior is exemplary studied with the
Barnase:Barstar model system.  In Fig.\ \ref{patch_size_plot}, the
encounter frequency has been obtained from simulations for
Barnase:Barstar-like model particles in the framework of \Mone\ at
several concentrations $c_0=\{5\mu$M$,125$nM$,2.5$nM$,125$pM$\}$ and
varying patch sizes $r$. All values in the figure have been scaled
with the concentration, which leads to data collapse. It is obvious
that as $r$ gets larger than $2R$ at around $r=40$\AA, the reactive
patch covers the whole model particle and we therefore cross over to
the Smoluchowski limit of isotropic reactivity, where
$k \sim r$. However, at high densities and large $r$, the
patches span a large part of the simulation box of edge length $L$,
and do immediately encounter for a threshold value of $r=r_{max}=L
\sqrt{3}/4$, where the sum of the patch diameters $4r$ equals the
triagonal. Thus, the encounter frequency must diverge with $\sim
1/(r_{max}-r)^\alpha$, where we suppose $\alpha=3$, as the volume of
configurational space without immediate encounter is decreasing with
$r^3$. This assumption in addition with the Smoluchowski behavior
would lead to $k\sim r/(r_{max}-r)^3$ for large $r$, which follows the
data in Fig.\ \ref{patch_size_plot} well (black dashed lines).

As already mentioned it is well known that the electrostatic interaction of
proteins can severely increase the association rate. However, under
physiological salt conditions, Coulombic interactions are screened by counter
ions in the solution on a small length scale of approximately
1 nm. Thus, deviations from case \Mone\ without effective charges will
only arise for small $r$. Fig.\ \ref{patch_size_electro_plot} shows the
results of respective simulations for \Mtwo\ compared to the results
for \Mone, as considered before. Indeed, for large patch radii $r$, the results
are similar, while for smaller $r$, the encounter rates in \Mtwo\ are clearly
higher compared to \Mone. However, the crossover to a power law behavior with
roughly $\sim r^{9/4}$ can be detected for very small $r$, but at a
prefactor of about 50 times larger than for \Mone.

\section{Discussion \label{summary_and_discussion}}

The main goal of this work was to model protein encounter in a generic
framework which allows us to include molecular
details without making future upscaling to larger complexes
impossible.  Our model approach incorporates steric, electrostatic and
thermal interactions of the proteins considered. These interactions
are thought to be the major factors governing protein encounter. Not
included are conformational changes of the proteins upon association,
related entropic terms, and the molecular nature of the surrounding
solvent that becomes relevant at close distances. The model parameters
are extracted from the atomic structures available in the protein data
bank by generally applicable protocols as described in Sect.\
\ref{model_and_methods}. In principle, these methods of data
extraction can be fully automatized.

The biggest advantage of our coarse-grained model is the possibility
to extend the simulations to large scales in terms of particle
numbers, time and system size. In many of the earlier studies
\cite{northrup_84, eltis_91, northrup_92, zhou_96}, the system was
prepared already close to encounter and the overall association rate
was then calculated via a sophisticated path-integral like procedure.
In contrast, our simulations account for the whole process of
diffusional encounter and is thus rather general, allowing for
spanning large time scales via our adaptive time step algorithm. In
particular, each set of simulations consists of $10^4$ to $10^5$ runs
of lengths up to the order of seconds and could be performed on a
standard CPU within hours of computer time.

Being able to directly obtain the first passage times (FPT) of the
encounter processes in our model allows to check the validity of
several phenomenological assumptions. First of all, the FPT
distribution matched very well a Poisson process with a single
stochastic rate, as seen in Fig.\ \ref{randinit_histogram}, which
validates the notions of encounter and association. Moreover, our
approach provides two ways of controlling the particle density and for
both cases the results corresponded well to the expected scaling.
First, the concentration is inversely correlated with the size of the
periodic boundary simulation box. We show that the encounter rate
grows linearly with the particle concentration. Furthermore, leaving
the box size constant, the concentration can also be varied by adding
a higher number $N$ of particles. Considering only the \emph{first}
encounter of any of the possible complementary pairs of model
particles, the mean first passage time to this event is not only
lowered by a factor $N$ but we show that the expected behavior is an
enhancement of the encounter frequency by $N^2$, which is nicely
matched by the results of the simulations. Therefore we can conclude
that the computational model studied here satisfies the general
requirement of stochastic bimolecular association processes that
describe binding by a single rate constant.

To test our model against known results we have chosen three
well-known bimolecular systems with different characteristics. The
Barnase:Barstar complex is the gold standard for protein-protein
association and characterized by relatively strong electrostatic
steering. The association of Cytochrome c and its peroxidase is even
more strongly affected by Coulombic attraction. Here, both proteins
have a rather spherical form. Finally, the p53:MDM2 complex has a
different characteristic with a very small net charge and a deep
cleft perfectly matching the small peptide p53, whose reactive
area is therefore nearly spanning over its whole surface. These model
systems were purposely chosen to check whether our effective
representations of the protein properties would lead to reasonable and
significantly distinguishable results. Indeed, this is the case as the
discussion of the results in Tab.\ \ref{main_data_table} in the
respective section shows.

When comparing the results for the encounter rates in Tab.\
\ref{main_data_table} with previous studies from the field of
bimolecular protein association, several aspects have to be kept in
mind. First, throughout this study, we do only consider the
\emph{encounter} of our model particles. As explained in the
beginning, the complete association of the complex still lacks the
step over a final free energy barrier, which is due to effects such as
the dehydration of the protein surfaces and thus requires more
detailed modelling. In the framework of our approach, this final step
could be modelled by a stochastic rate criterion, where the rate can
be obtained by transition state theory from the energy landscapes
characterized in atomistic calculations. In any case, any additional
process to be included can only lower the values found in our study.

In the work on Barnase:Barstar by Schreiber et al.
\cite{schreiber_96}, the authors reported that the association between
Barnase and Barstar is a diffusion-limited reaction. The argument for
this is that the association rates at high ionic concentrations in the
solution, i.e.\ for the limit in which the electrostatic steering gets
negligible, are clearly lowered by the addition of glycerol, which
will lead to slower diffusion.  Assuming diffusion control, the
reactive step over the final barrier should be kinetically
unimportant, as generally discussed in Ref.\ \cite{bell_78}. Indeed,
we see that our results for the encounter rates lead to values in the
correct order of magnitude of $k\approx10^9$M$^{-1}$s$^{-1}$, which is
similar to the experimental value obtained by Schreiber et al.\ for
the association constant of Barnase:Barstar
$k=8\cdot10^8$M$^{-1}$s$^{-1}$ at physiological salt
\cite{schreiber_93}. However, the basal association rate, i.e.\ the
rate at high ionic strength, is reported as
$k<10^6$M$^{-1}$s$^{-1}$ from experiments
\cite{schreiber_96}. Given that the association process of Brn:Brs is
diffusion limited, these findings should actually coincide with our
values for \Mone. But as we already discussed in the results section,
in our simulations the influence of the effective electrostatics
introduced in \Mtwo\ do not result in such a drastic change of the
encounter behavior.

In several earlier approaches, similar problems have been addressed by
computational and analytical studies. In work by Zhou and coworkers, basal
encounter rates for particles with reactive patches have been found to be
$k=4\cdot10^6$M$^{-1}$s$^{-1}$ \cite{zhou_93} and
$k=10^7$M$^{-1}$s$^{-1}$ \cite{zhou_97}, that is closer to the basal
rates reported by Schreiber and coworkers. It has to be noted that,
in both cases, the patches were flat areas above the surface of the spherical
model particles, which had a smaller angular extension compared to our cases,
and especially required a much closer translational approach ($0.7$\AA\ in
\cite{zhou_93}) to form the encounter. If we expand the graph in Fig.\
\ref{patch_size_plot} to smaller patch radii like $r=1$\AA, we also find basal
rates in the order of $k=10^7$M$^{-1}$s$^{-1}$. Also, the deviation between
\Mone\ and \Mtwo, i.e.\ the influence of the effective electrostatics, is more
prominent and could enhance the encounter rate by about two orders of
magnitude, which is consistent with the findings in the previously cited
work. There, the effect of Coulombic interaction is reflected with a Boltzmann
factor due to a pairwise Coulomb energy. This approach works well, as shown in
Ref.\ \cite{zhou_96}, and has been recently used in a more complex model study
of the energy landscape of protein-protein association \cite{ramzi_07_a,
ramzi_07_b}.

Any model for the reaction patches has to rely on results obtained
from more detailed modeling. The surface of a protein is typically
densely covered by water molecules due to the hydrophilic nature of
its surface. This hydration shell has a thickness of about 3\AA\ and
will therefore in principal hinder the approach of two proteins to
distances below 6\AA. Setting the encounter patches to values below
this threshold of 3\AA\ would then mean that part of the dehydration
would already have happened before the encounter is actually formed,
which is probably hardly described by simple diffusion with drift.
Moreover, all of the considered protein
systems feature distinct key-lock binding interfaces regarding the
steric structure, apart from some flexibility due to intrinsic thermal
motion. Therefore, it makes sense to represent the encounter area by a
three-dimensional extended object rather than by a flat surface region.
Indeed, the results of our studies show that our approach 
is capable of reproducing encounter rates in a reasonable order
of magnitude, qualitatively reproducing generally expected features. In
an in-depth investigation of the dependency of the encounter rate and
the patch radius it is shown, that the choice of the geometry of the
reactive area is at least as crucial for the results as definition of
the model interactions and its parameters. In principal, one could
think of the patch radius as a valuable tuning parameter to fit
experimental results and the encounter kinetics in the computational
model. 

Our approach makes it possible to observe general features of the
encounter process. In particular, we dissect the pathway to the encounter
complex in several levels of alignment between our model proteins. As we
observe the full trajectory to encounter in our simulations, we are able to
extract the number of unsuccessful contacts $N$ between the proteins
until they finally reach a reasonably aligned state to bind. The distribution
of $N$ is again in all cases well described by a single exponential decay.
This behavior is not obvious as the probability of success for one contact is
depending on several aspects of diffusion in a complex manner. First, the
closer the rotational alignment at the beginning of the contact is to the
encounter state, the higher is the probability of success. Second, this
initial alignment is also coupled to the last contact if the time in between
$T_\text{off}$ is small. Finally, longer contact resting times $T_\text{on}$
also increase the probability of encounter. It is interesting, that all these
effects still lead to a simple Poisson distribution of the number of contacts
$N$ when averaging over the initial conditions as it is done in this work.
Furthermore, we find that the distributions of these resting and return times
cannot be described by a Poisson process, but are consistent with the
expectations for a spatially constricted random walk in three dimensions. We
find that the particular mean FPT to encounter is in most of the cases
directly proportional to the number of unsuccessful contacts. This seems to be
a very fundamental qualitative feature irrespective of the details of the
proteins and the applied model. However, for Cytc:CCP the behavior is
qualitatively different, which is consistent with earlier studies of this
highly electrostatically steered complex.

In summary, here we have presented a Langevin equation approach to
protein-protein association which in principle allows us to combine
long simulation times and large systems with molecular details of the
involved proteins. This first study has focused on bimolecular
reactions and has proven that this approach is capable of reproducing
known association rates with a reasonable dependance on the main
parameters involved. One special strength of our approach is that it
allows us to address the details of the binding pathway, for example
by measuring the statistics of unsuccessful contacts before encounter.
In the future, this approach will be extended to large protein
complexes of special biological interest.

\begin{acknowledgments}
  We thank Christian Korn for many helpful discussions. This work was
  supported by the Volkswagen Foundation through grants I/80469 and
  I/80470 to V.H. and U.S.S., respectively. J.S. and U.S.S. are
  supported by the Center for Modelling and Simulation in the
  Biosciences (BIOMS) at Heidelberg and by the Karlsruhe Institute of
  Technology (KIT) through its Concept for the Future.
\end{acknowledgments}

\clearpage

\begin{table}
\caption{\label{parameter_table} 
Protein structures and parameters used in the study. The coordinates of the patches (${\mathbf r}_\text{patch}$) and the
dipole moment ($\mathbf p$) are given relative to the center of mass.
}
\begin{tabular}{c c c c c c c c}
\hline\hline
Protein & System        & PDB code   & Ref.\ & $R_\text{gyr}$/\AA\ & $q$/$e$ & ${\mathbf r}_\text{patch}$/\AA\ & $\mathbf p$/\AA\ $e$ \\ 
\hline
Barnase & $\mathcal S1$ & {\tt 1brs} & \cite{buckle_94, gabdoulline_97} & 14.68 & $2  $ & $(5.43,-4.75,	-3.41)$ & $(3.84,-0.67,	-36.15 )$ \\
Barstar &               &            &                                  & 13.42 & $-6 $ & $(-6.05,3.75,	6.34 )$ & $(54.75,-14.39,0.621  )$ \\
Cytc    & $\mathcal S2$ & {\tt 2PCC} & \cite{pelletier_92}              & 13.89 & $6  $ & $(-2.08,7.99,	-3.83)$ & $(-7.03,117.79,-10.99 )$ \\
CCP     &               &            &                                  & 20.00 & $-13$ & $(8.89,-8.19,	11.46)$ & $(-23.47,93.85,-171.82)$ \\
p53     & $\mathcal S3$ & {\tt 1YCR} & \cite{kussie_96}                 & 10.20 & $-2 $ & $(0.28,0.21,	0.59 )$ & $(21.64,-17.73,-17.12 )$ \\
MDM2    &               &            &                                  & 16.81 & $1  $ & $(1.577,-4.74,-0.51)$ & $(75.07,14.51,4.67   )$ \\
\hline \hline
\end{tabular}
\end{table}

\clearpage

\begin{table}
  \caption{\label{main_data_table} Encounter rates $k$ which have been averaged over several simulations at
    different concentrations as given in the text. The values are given in $k$
    / $10^9$M$^{-1}$s$^{-1}$ for the three different versions of our model.
    $\mean{N}$ are average values for the number of unsuccessful
    contacts before encounter. $\mean{N}$ is basically independent of the
    concentration. Therefore it is again averaged over the different
    simulations for each of the chosen systems. The errors were determined by
    one standard deviation from the 8 values obtained at different
    concentrations. Some of the choices for the patch radius were not
    applicable to \Mthree, as for these cases an encounter was completely
    prevented by the detailed excluded volume model.}
  \begin{tabular}{c c@{\hspace{1mm}} | @{\hspace{1mm}}rcl @{\hspace{3mm}}rcl @{\hspace{3mm}}rcl@{\hspace{1mm}} |@{\hspace{1mm}} rcl@{\hspace{3mm}} rcl@{\hspace{3mm}} rcl }
    \hline\hline
    System & Patch radius & \multicolumn{3}{c}{$k$(\Mone)} & \multicolumn{3}{c}{$k$(\Mtwo)} & \multicolumn{3}{c}{$k$(\Mthree)} & \multicolumn{3}{c}{$\mean{N}$(\Mone)} & \multicolumn{3}{c}{$\mean{N}$(\Mtwo)} & \multicolumn{3}{c}{$\mean{N}$(\Mthree)} \\ 
    \hline
    Brn:Brs  &   10.0    & $1.56$&$\pm$&$0.04 $ & $2.76$&$ \pm$&$ 0.07$ & $2.02$&$ \pm$&$ 0.02$ & $474 $&$\pm$&$ 2$ & $198$&$\pm$&$ 5 $ & $282$&$\pm$&$10 $  \\
             &    6.0    & $0.57$&$\pm$&$0.01 $ & $2.13$&$ \pm$&$ 0.01$ & $1.34$&$ \pm$&$ 0.08$ & $1140$&$\pm$&$4 $ & $232$&$\pm$&$ 8 $ & $534$&$\pm$&$50 $  \\
             &    3.0    & $0.13$&$\pm$&$0.001$ & $1.28$&$ \pm$&$ 0.03$ & $    $&$  - $&$     $ & $4120$&$\pm$&$10$ & $653$&$\pm$&$15 $ & $   $&$ - $&$   $  \\
    \hline
    Cytc:CCP &   10.0    & $1.12$&$\pm$&$ 0.02 $ & $4.31$&$ \pm$&$ 0.20$ & $4.15$&$ \pm$&$ 0.15$ & $842 $&$\pm$&$ 3$ & $61 $&$\pm$&$ 5 $ & $71  $&$\pm$&$ 7 $  \\
             &    6.0    & $0.40$&$\pm$&$ 0.01 $ & $4.29$&$ \pm$&$ 0.21$ & $4.05$&$ \pm$&$ 0.09$ & $2040$&$\pm$&$20$ & $30 $&$\pm$&$ 3 $ & $77  $&$\pm$&$10 $  \\
             &    3.0    & $0.09$&$\pm$&$ 0.001$ & $4.03$&$ \pm$&$ 0.05$ & $0.21$&$ \pm$&$ 0.02$ & $7540$&$\pm$&$63$ & $21 $&$\pm$&$ 3 $ & $4160$&$\pm$&$375$ \\
    \hline
    p53:MDM2 &   10.0    & $2.05$&$\pm$&$ 0.05 $ & $2.51$&$ \pm$&$ 0.04$ & $1.27$&$ \pm$&$ 0.02$ & $362 $&$\pm$&$ 2 $ & $266$&$\pm$&$ 4 $ & $823 $&$\pm$&$10  $  \\
             &    6.0    & $0.80$&$\pm$&$ 0.01 $ & $1.12$&$ \pm$&$ 0.01$ & $0.15$&$ \pm$&$ 0.01$ & $815 $&$\pm$&$10 $ & $582$&$\pm$&$13 $ & $8720$&$\pm$&$200 $  \\
             &    3.0    & $0.19$&$\pm$&$ 0.002$ & $0.28$&$ \pm$&$ 0.01$ & $    $&$  - $&$     $ & $2900$&$\pm$&$35$ & $2550$&$\pm$&$30$ & $     $&$-  $&$ $  \\
    \hline \hline
  \end{tabular}
\end{table}

\clearpage

\begin{figure}
  \caption{ Scheme to visualize the different variants of the model for the
    three considered model systems. The color code is: yellow for Barnase,
    cytochrome c and p53; green for Barstar, cytochrome c peroxidase and
    MDM2. The respective reaction patches are shown in white. \Mone\ only
    includes a simple steric interaction. \Mtwo\ has an additional effective
    electrostatic interaction, here denoted with red arrows showing the
    direction of the dipole of the model particles. In \Mthree, the excluded
    volume is modeled in more detail as a collection of smaller beads. The
    transparent blue spherical surface marks the volume used in \Mone\ and
    \Mtwo\ for the sake of comparison. Finally, the bottom row shows
    surface representations of the atomistic structures taken from the protein
    database.
    \label{different_model_scheme}
  }
\end{figure}

\begin{figure}
  \caption{Logarithmic plot of the distribution of the first passage time to encounter $T$
    between a single pair of Barnase and Barstar model particles in a cubic
    boundary box of edge length $L=2370$\AA, representing a concentration of
    $0.125\mu$M for each protein. The dashed line represents a single
    exponential fit to the data points, which shows the expected behavior with
    respect to the encounter frequency $k=\mean{T}^{-1}$.
    \label{randinit_histogram}
  }
\end{figure}

\begin{figure}
  \caption{Simulated encounter frequencies for a single pair of Barnase and
    Barstar model particles in cubic boundary boxes of different sizes
    representing different concentrations. The dashed line is a linear fit to
    the data.
    \label{randinit_concentrations}
  }
\end{figure}

\begin{figure}[ht]
  \caption{ Encounter frequency for different number of Barnase:Barstar pairs
    leaving the size of the boundary box constant. The red data points show
    the encounter frequencies as obtained from simulations, while the green
    line represents the function $CN^2$, where $N$ is the number of particle
    pairs and $C$ is a fitted prefactor.
    \label{randinit_finite_size}
  }
\end{figure}

\begin{figure}
  \caption{ Different alignment states during the encounter process. \Aone\
    proteins are completely unaligned. In state \Atwo, referred to as
    \emph{contact} in this paper, the proteins are translationally aligned,
    i.e.\ they are close enough to actually encounter (denoted by the overlap
    of the lightened area around the model particles), but lack the correct
    orientation. \Athree\ proteins reached the encounter meaning that the
    reactive patches are in translational and rotational alignment.
    \label{alignment_states_scheme}
  }
\end{figure}

\begin{figure}
  \caption{ Logarithmically plotted distribution of the number of approaches $N$
    between a Barnase and a Barstar particle with incorrect rotational
    alignment. The dashed line is an exponential fit to the data. 
    \label{alig_no_example}
  }
\end{figure}

\begin{figure}
  \caption{ Double-logarithmic plot of the distribution of resting and return
    times of the translationally aligned state (\Atwo\ in Fig.\
    \ref{alignment_states_scheme}).
    \label{alig_onoff_example}
  }
\end{figure}

\begin{figure}
  \caption{Correlation plot of encounter rate $k$ and mean number of contacts
    $\mean{N}$ with all the data from Tab.\ \ref{main_data_table}.
    \label{correlation_plot}
  }
\end{figure}

\begin{figure}
  \caption{Encounter rates in dependency of the patch size for the
    Barnase:Barstar model system in the \Mone\ variant.
    \label{patch_size_plot}
  }
\end{figure}

\begin{figure}
  \caption{Comparison of \Mone\ and \Mtwo\ similar to Fig.\
    \ref{patch_size_plot} for small patch sizes. For larger patch sizes there is
    no substantial difference.
    \label{patch_size_electro_plot}
  }
\end{figure}

\clearpage

\includegraphics[width=\textwidth]{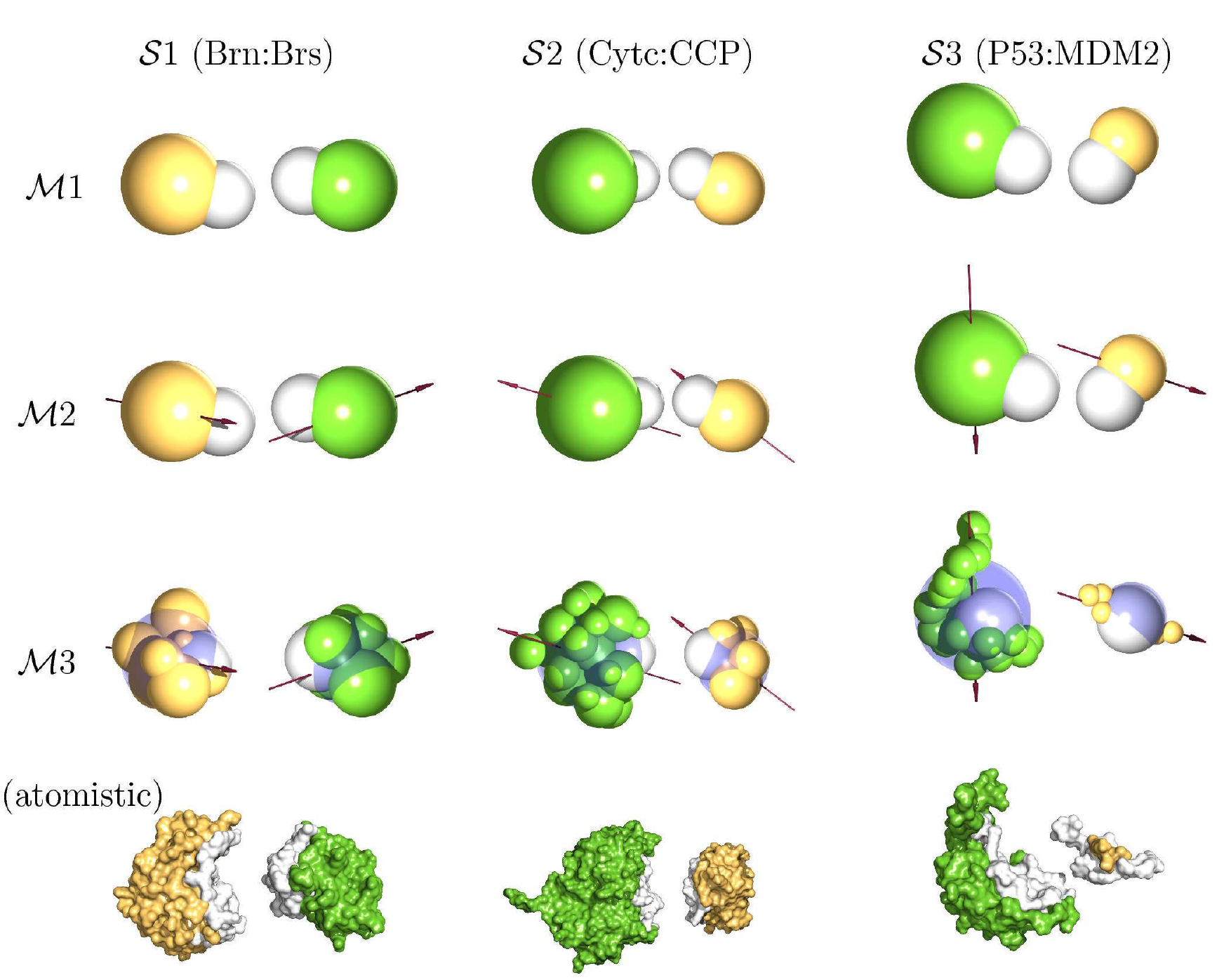}

\fig{different_model_scheme}

\clearpage

\includegraphics[width=\textwidth]{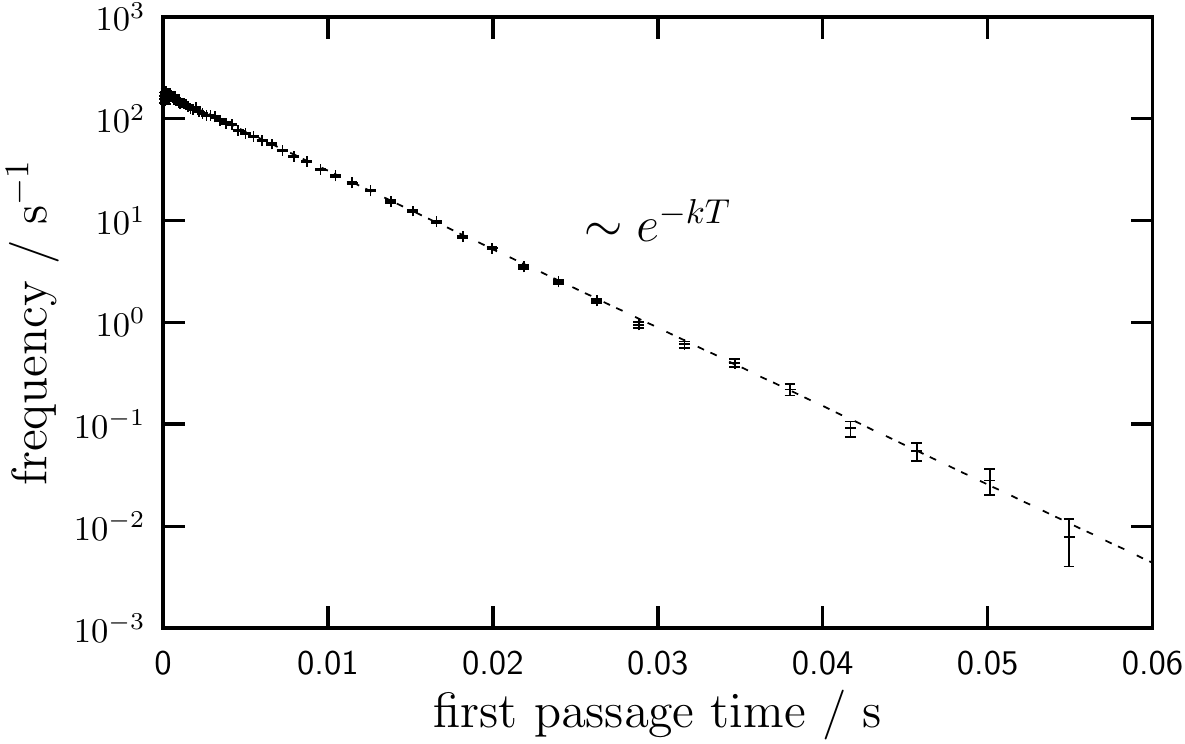}

\fig{randinit_histogram}

\clearpage

\includegraphics[width=\textwidth]{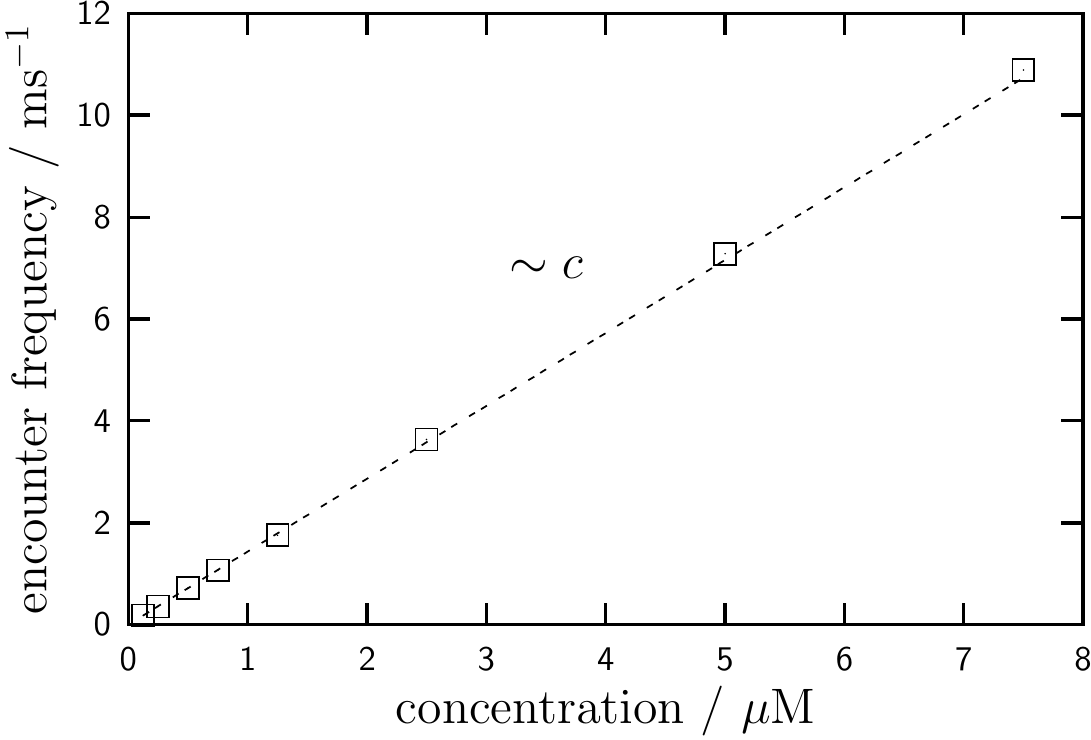}

\fig{randinit_concentrations}

\clearpage

\includegraphics[width=\textwidth]{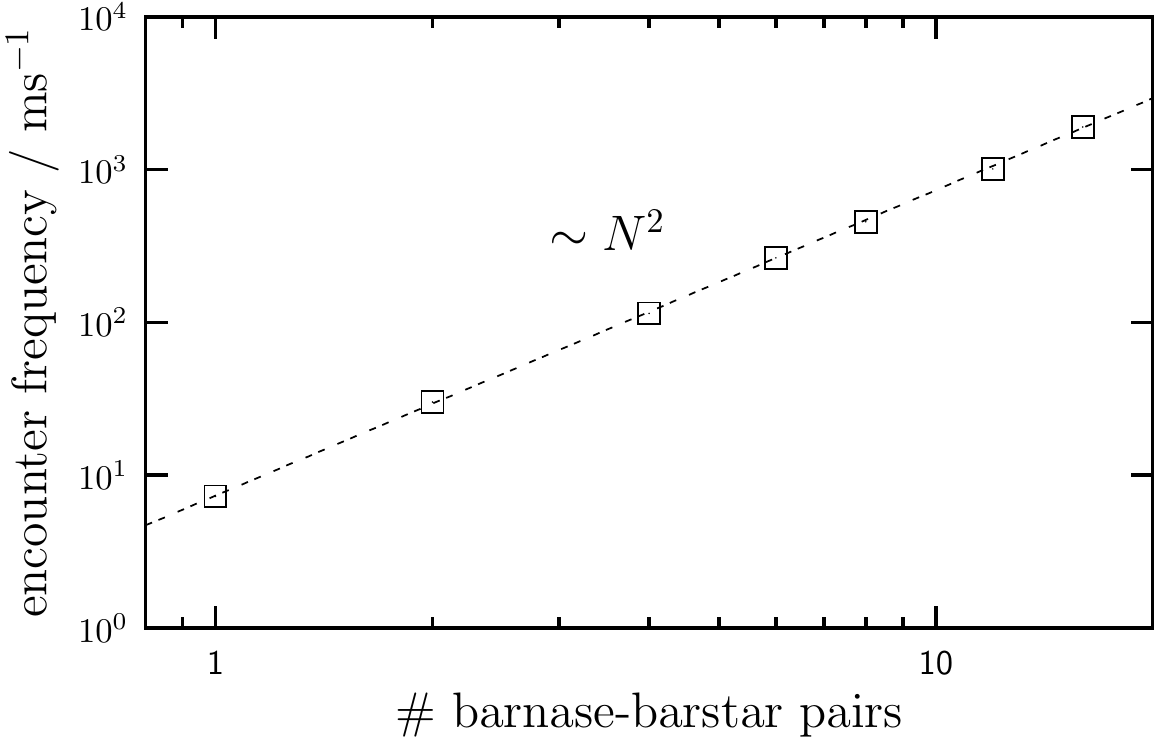}

\fig{randinit_finite_size}

\clearpage

\includegraphics[width=\textwidth]{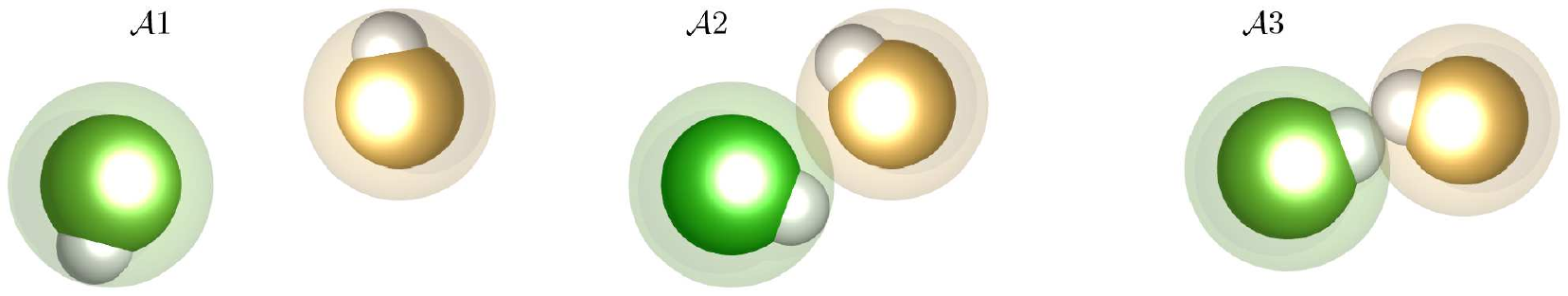}

\fig{alignment_states_scheme}

\clearpage

\includegraphics[width=\textwidth]{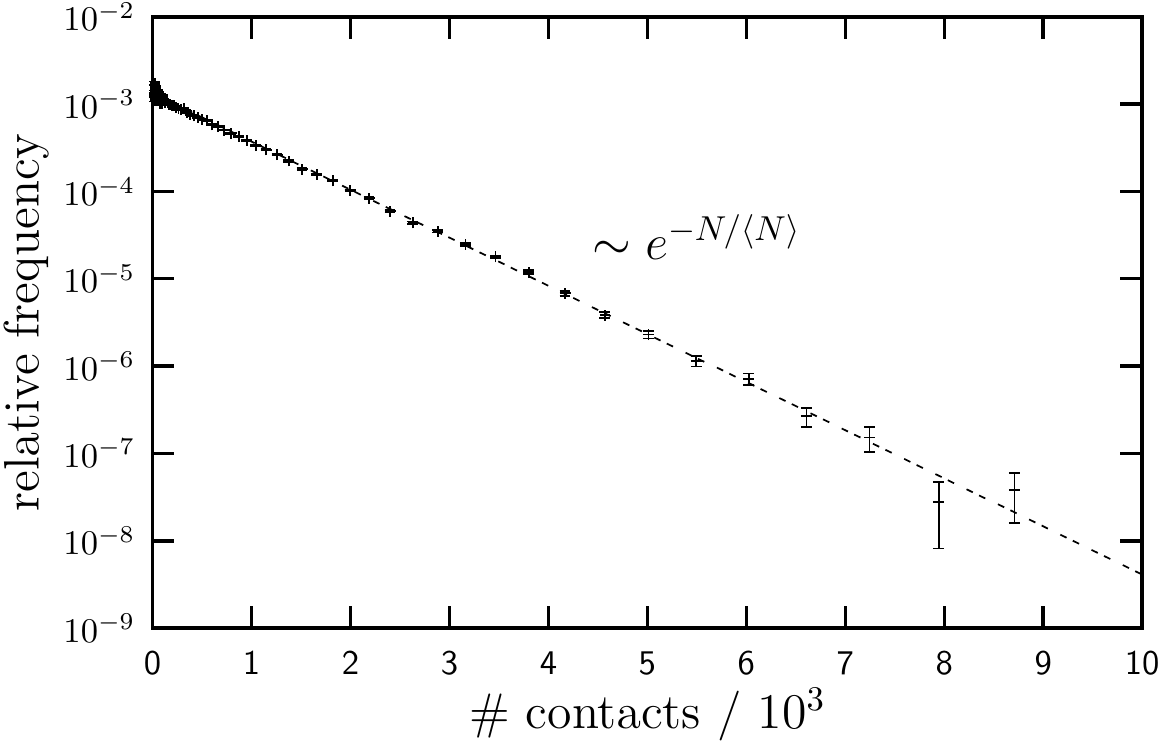}

\fig{alig_no_example}

\clearpage

\includegraphics[width=\textwidth]{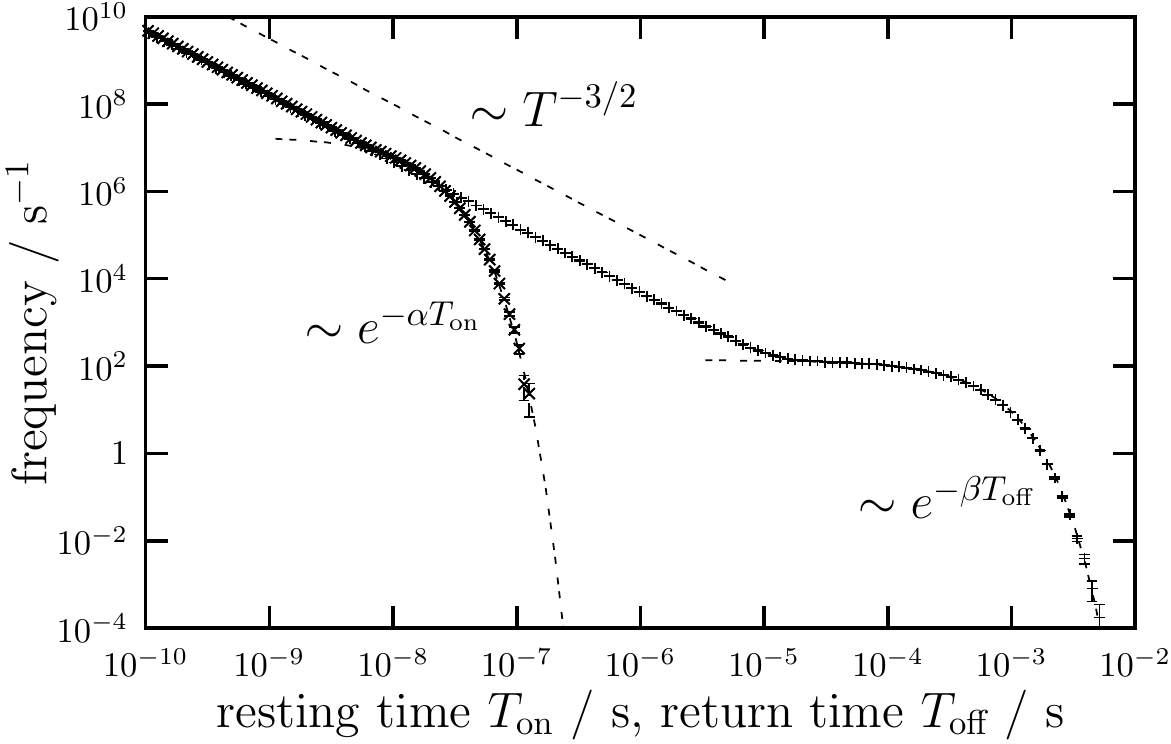}

\fig{alig_onoff_example}

\clearpage

\includegraphics[width=\textwidth]{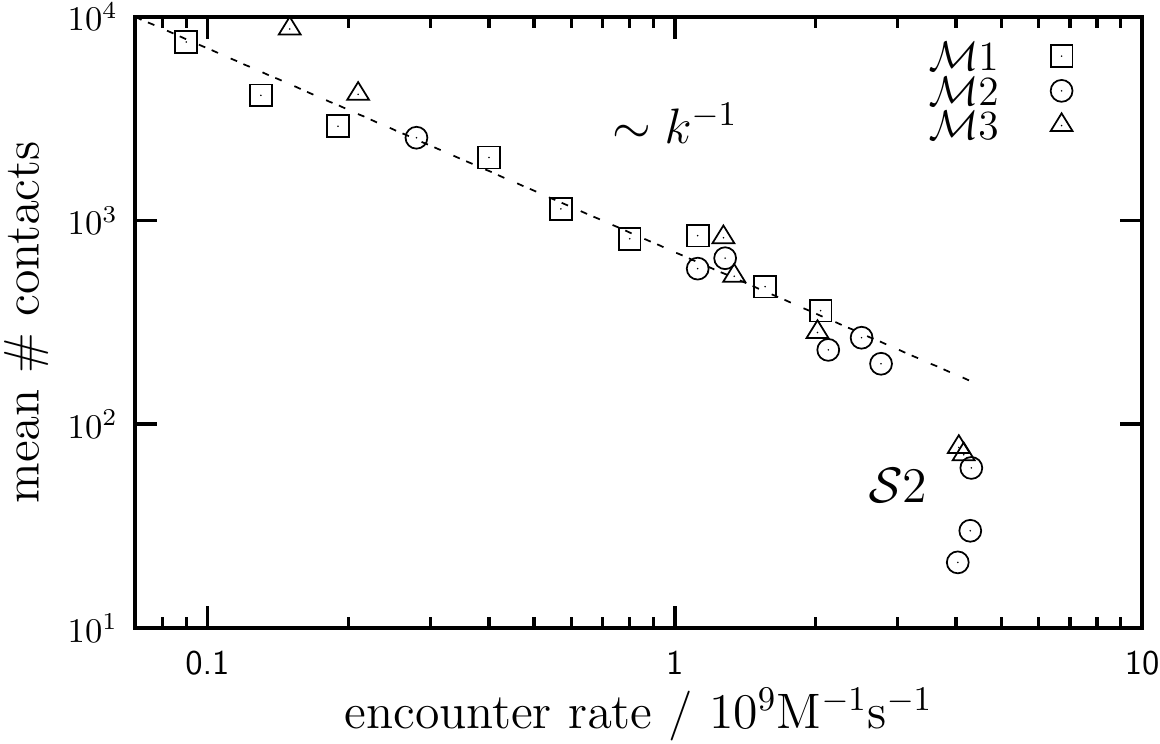}

\fig{correlation_plot}

\clearpage

\includegraphics[width=\textwidth]{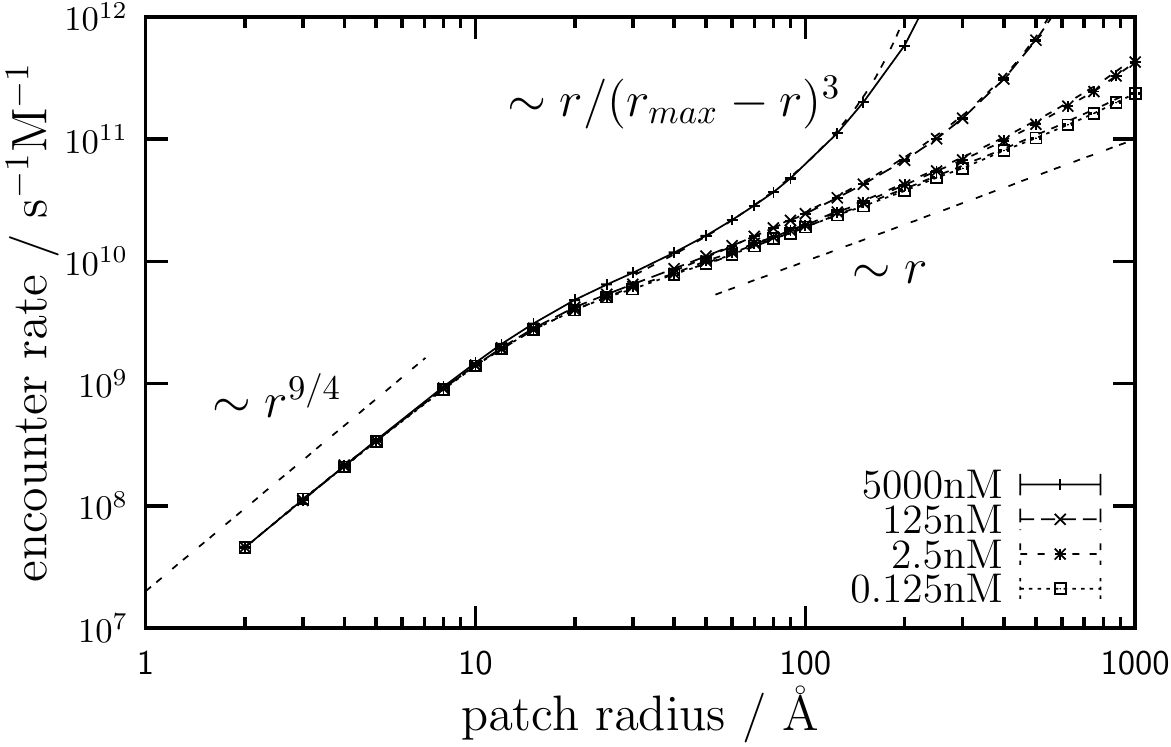}

\fig{patch_size_plot}

\clearpage

\includegraphics[width=\textwidth]{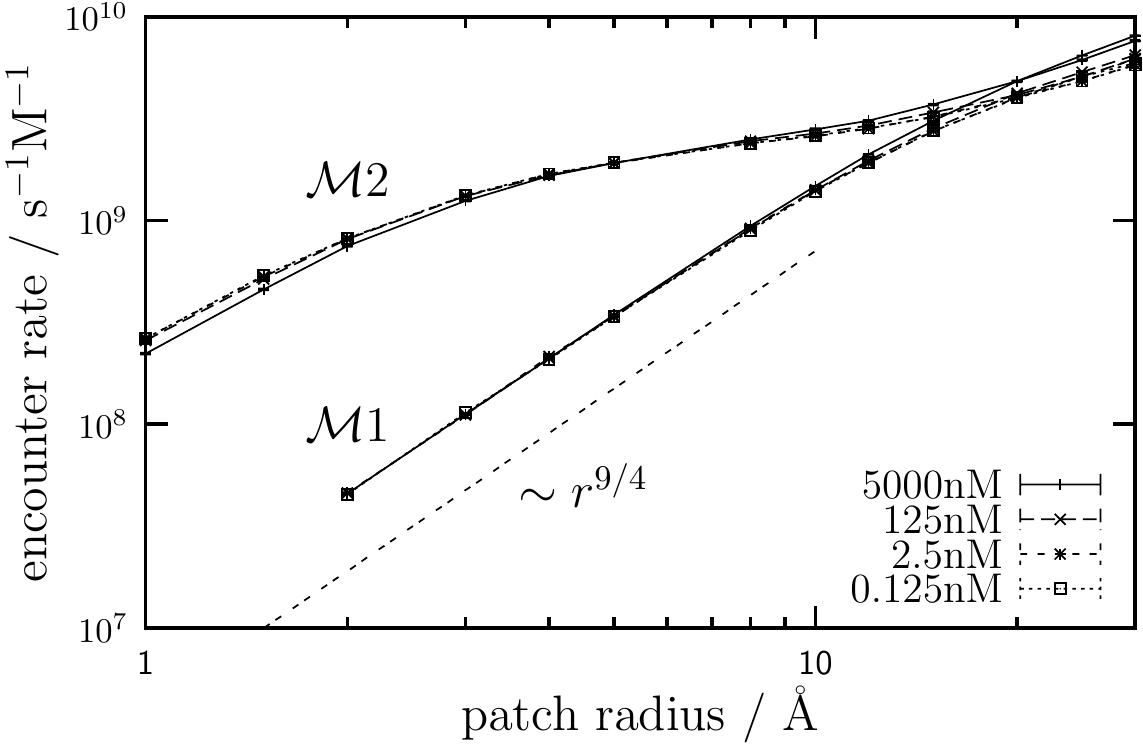}

\fig{patch_size_electro_plot}

\end{document}